# INTRODUCTION TO COMPUTER ANIMATION AND ITS POSSIBLE EDUCATIONAL APPLICATIONS☆

Sajid Musa [a], Rushan Ziatdinov [b*], Carol Griffiths [c]

[a,b]Department of Computer and Instructional Technologies, Fatih University, 34500 Buyukcekmece, Istanbul, Turkey
E-mail: sajidmusa004@gmail.com and rushanziatdinov@gmail.com
[c]Department of Foreign Language Education, Fatih University, 34500 Buyukcekmece, Istanbul, Turkey
E-mail: carolgriffiths5@gmail.com

**Abstract**

Animation, which is basically a form of pictorial presentation, has become the most prominent feature of technology-based learning environments. It refers to simulated motion pictures showing movement of drawn objects. Recently, educational computer animation has turned out to be one of the most elegant tools for presenting multimedia materials for learners, and its significance in helping to understand and remember information has greatly increased since the advent of powerful graphics-oriented computers.

In this book chapter we introduce and discuss the history of computer animation, its well-known fundamental principles and some educational applications. It is however still debatable if truly educational computer animations help in learning, as the research on whether animation aids learners' understanding of dynamic phenomena has come up with positive, negative and neutral results.

We have tried to provide as much detailed information on computer animation as we could, and we hope that this book chapter will be useful for students who study computer science, computer-assisted education or some other courses connected with contemporary education, as well as researchers who conduct their research in the field of computer animation.

## I. Introduction

For the past two decades, the most prominent feature of the technology-based learning environment has become animation (Dunbar, 1993). Mayer and Moreno (2002) state that animation is a form of pictorial presentation - a definition which also refers to computer-generated motion pictures showing associations between drawn figures. Things which correspond to this idea are: motion, picture and simulation. As far as videos and illustrations are concerned, these are motion pictures depicting movement of real objects.

---

The birth of pictorial forms of teaching has been observed to have developed as a counterpart to verbal forms of teaching (Lowe, 2004; Lasseter et al., 2000; Mosenthal, 2000). Although verbal ways of presentation have long dominated education, the addition of visual forms of presentation has enhanced students' understanding (Mayer, 1999; Sweller, 1999). In fact, some disciplines are taught in universities which deal with dynamic subject matter, and animation or graphic illustration is more favoured as a way of addressing the difficulties which arise when presenting such matters verbally or numerically (Lowe, 2004).

Even though such multimedia instructional environments hold potential for enhancing people's way of learning (Lowe, 2004; Lasseter et al., 2000; Mosenthal, 2000) there is still much debate surrounding this area; indeed animation presentations are less useful for the purposes of education and training than was expected. Moreover, little is known about the way animation needs to be designed in order to aid learning (Plötzner & Lowe, 2004) and not to act solely as a way to gain aesthetic attraction. For instance, some animators who work in the entertainment industry create animations for the sake of entertainment and they are therefore unlikely to be interested in helping to build coherent understanding using their work (Lowe, 2004).

In some cases, animation can even hold back rather than improve learning (Campbell et al., 2005), and may even not promote learning depending on how they are used (Mayer & Moreno, 2002). Animation may possibly require greater cognitive processing demands than static visuals as the information changes frequently, especially critical objects, and thus cognitive connection can be lost during the animation (Hasler et al., 2007).

As noted by Hegarty (2005) in *Learning with Animation: Research Implications for Design*, "the current emphasis on ways of improving animations implicitly assumes a bottom-up model animation comprehension… Comprehension is primarily a process of encoding the information in the external display, so that improving that display necessarily improves understanding." Similarly, Lowe noted in his work *Learning from Animation Where to Look, When to Look,* that the main problem that the developers of multimedia learning materials face is the lack of principled guidance on how some elements of such materials should be designed in order to enable comprehension.

Mayer and Moreno (2000) examined the role of animation in multimedia learning; they also presented a cognitive theory of multimedia learning and were able to summarize the programme of their research. They come up with seven principles for the use of animation in multimedia instruction. Some of these principles were multimedia principles; students learn more deeply when narration and animation come together than narration or animation alone. Learners can easily create mental connections between corresponding words and pictures when both animation and narration are presented. The other principle was the coherence principle; they say that students learn more deeply from both animation and narration especially when irrelevant words, sounds (even music) and clips are not present. This is due to the chances of the learner experiencing difficulty in building mental connections because of fewer cognitive resources between relevant portions of the narration and animations (Lowe, 2004).

Hasler (2007) investigated the effect of learner controlled progress in educational animation on instructional efficiency. Based on her paper, three audio-visual computer animations and narration-only presentations were used to teach primary school students the determinants of day and night. One of the animations was system-paced using an uninterrupted animation. The results of the experiment showed that the group which had a two learner paced groups displayed higher test performance compared with the other two (Hasler, 2007). Table 1 provides an overview on the works done in this field.

**Table 1.** Brief comparison of the previous work done in the field of educational animation.

| Author/Year | Paper Title | Aim of the Paper | Concise View of the Study (Findings/Problems) | Conclusion |
|---|---|---|---|---|
| Mayner & Moreno (2002) | Animation as an aid to Multimedia Learning. | Examine the role of animation in multimedia learning. | *Multimedia instructional messages* and *micro-worlds* were defined; giving rise to argument on "How should animation be presented to promote understanding of multimedia explanation?"; Took *Information Delivery Theory* into account; Presented a *Cognitive Theory of Multimedia Learning* comparisons and computations were made based on the results given by the college students from the tests and experiments they participated in resulting in a table of Seven Principles of Multimedia Learning. | There are seven principles for the use of animation in multimedia instruction; The *Cognitive Theory of Multimedia Learning* is more consistent than the *Information Delivery Theory*. |
| Lowe (2004) | Animation and Learning: Value for Money? | To show the effectiveness of animation's potential to play a role in cognitive function. | Two assumptions about the role of animation in education: (affective function & cognitive function); There are two distinct types of animation problem: (overwhelming & underwhelming); Stated that today's educational animations are called "behaviorally realistic" depictions. | Animation needs to be aesthetically made and supported if it is to achieve educational potential; A systematic approach and progress beyond current adherence and behavioural realism should be observed. |
| Hasler et al., (2007) | Learner Control, Cognitive Load and Instructional Animation | Examines the influence of two different forms of learner-controlled pacing during a temporary, audio visual animation on instructional efficiency. | Cognitive Load theory (CLT) gives a model for instructional design depending on human cognitive architecture; The effectiveness of instructional activities are determined from cognitive architecture; Pertaining to problems linked with high unnecessary load | The outcomes gained provided an experimental aid for the soundness and success of application of the principles derived from (CLT) in the context of multimedia learning. |

| | | | because of processing transient information, learners manage the pace and segmentation of presentation (instruction) into meaningful segments; Researchers ended-up with positive effects for learner-controlled pacing and segmentation while others found the other way around. | |
|---|---|---|---|---|

*Organization*

The rest of this book chapter is organized as follows. In Section II we briefly review the chronological order of the history of animation. In Section III we discuss the fundamental principles of computer animation. In Section IV we discuss the usage of computer animation in education. In Section V we talk about the future of computer animation. In Section VI we conclude our paper and suggest future work entitled *Theoretical Aspects of Creating Educational Computer Animation based on the Psychological Characteristics of the Human Temperament*.

**II. History of Animation**

Looking at the past and the present, animation has evolved over time. It started with pieces of paper and rope in 1828 and is today 3D animation videos. In this section, we will list the chronological order of the development of animation and animation devices which have evolved and improved over the past two centuries. We have also included the most famous animation characters in the history of animation.

- **Thaumatrope**

A thaumatrope (invented by Paul Roget in 1828) is a simple mechanical toy which creates the illusion of movement. Thaumatrope means "wonder turner" derived from the Greek words: θαῦμα "wonder" and τρόπος "turn". Roget was the first person in history to create such a device which produces the illusion of movement. In order to enjoy this animation, one would only need one small round piece of paper with pictures on it and thin ropes fixed at both ends of the shape (Figure 1). Below shows what a thaumatrope (Figure 1) is and how the illusion of movement is produced (Figure 2).

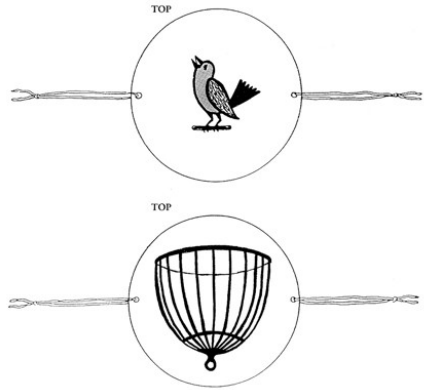
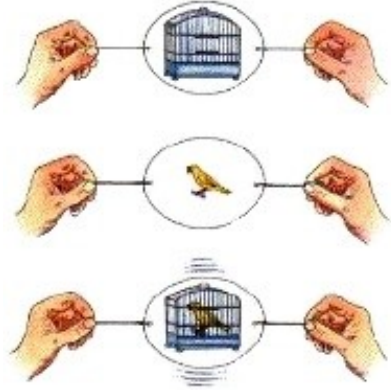

Figure 1. Thaumatrope.

Figure 2. Illusion of movement.

- **Phenakistoscope**

After the invention of the thaumatrope, the phenakistoscope followed made possible by J.A. Ferdinand Plateau in 1832. This device uses the persistence of vision principle to create an illusion of movement. Phenakistoscope originated from the Greek φενακίζειν (phenakizein), meaning "to trick or cheat"; as it tricks the eye by making the figures in the pictures appear to move. It is composed of six similar images in different positions taken in order to relay the movement. A very simple example is a running reindeer and jumping frogs (Figure 3).

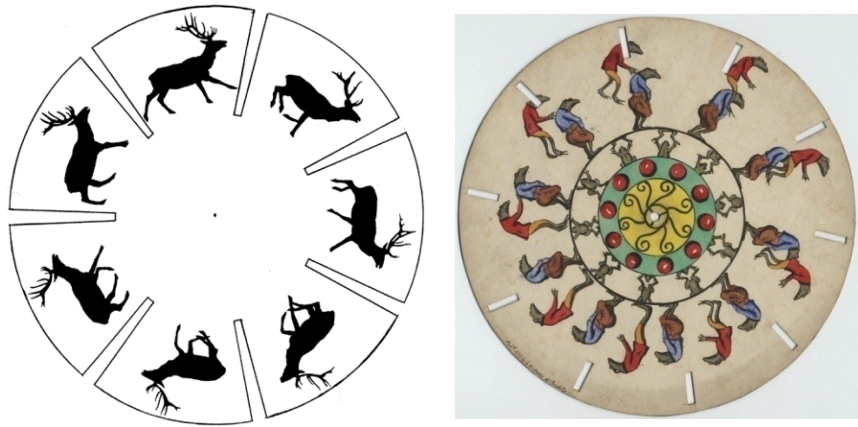

Figure 3. Examples of moving images using a phenakistoscope.

- **Zoetrope**

In 1843, William Horner, a British mathematician invented the zoetrope. A zoetrope produces an illusion of movement from a rapid succession of static pictures. Derived from the Greek words ζωή *zoe*, "life" and τρόπος *tropos*, "turn" this forms a "wheel of life".

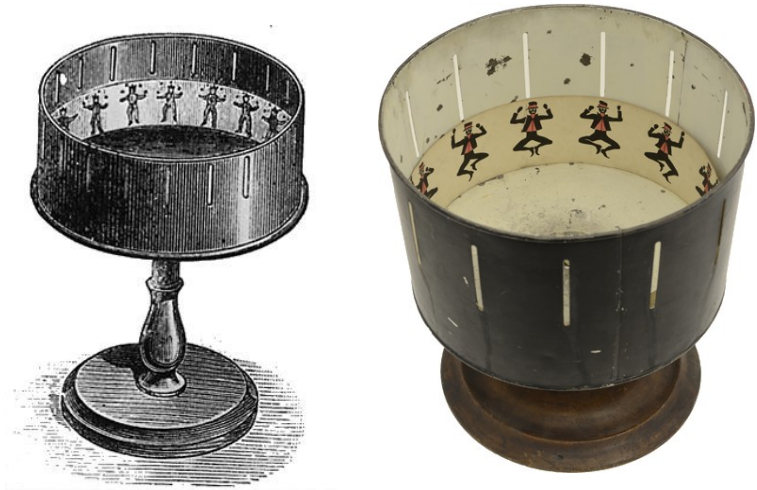

Figure 4. Different versions of the zoetrope.

- **Praxinoscope**

Almost the same as a zoetrope, the only difference was the integration of a mirror to the device which makes the viewer more comfortable as they watch the movement of the objects. It was designed by Emile Reynaud in 1877 and was known as the "action viewer".

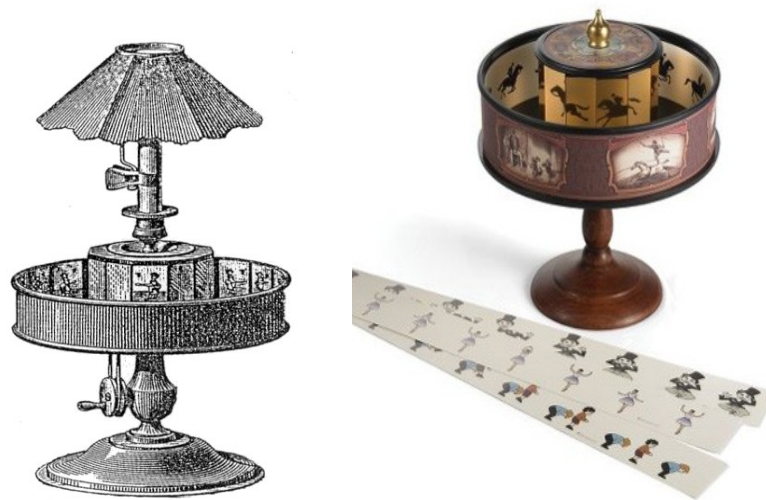

Figure 5. Praxinoscope in black and white, and a version in colour.

- **Kinestoscope**

An early motion picture exhibition device was invented in 1888 by Thomas Edison together with his colleague Eadweard Muybridge. The kinestoscope was designed for films to be viewed through the window of a cabinet (Figure 6). Kinestoscope means the "view of movement" from the Greek words *κίνησις* "movement" and *σκοπός* "movement".

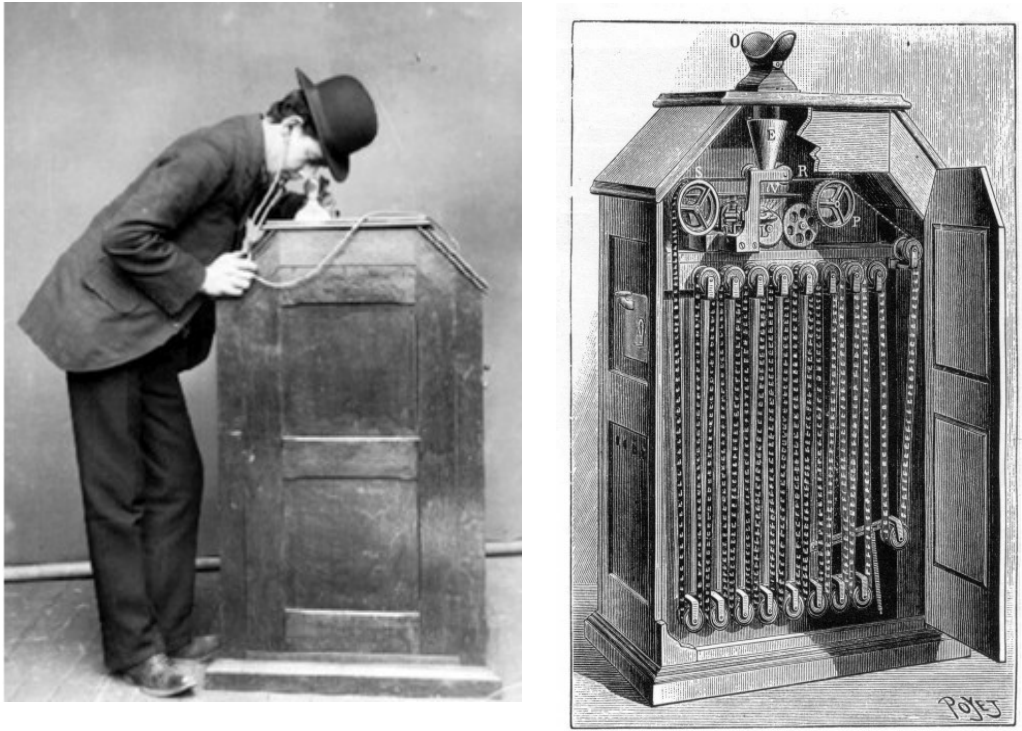

Figure 6. Man views the motion picture exhibit in a kinestoscope.

- **Multiplane Camera and Storyboard**

Walt Disney and his colleagues had a problem with creating realistic animation and how to conserve time while creating it. Then they came up with a great solution which can be considered another innovation in the field of animation - the multiplane camera (Figure 12). The multiplane camera is a piece of equipment designed to make cartoons more realistic and enjoyable. It uses stacked panes of glass each with different elements of the animation (Figure 13). With this, it allowed for the reuse of backgrounds, foregrounds, or any elements not in motion. The multiplane camera was developed by a Walt Disney Productions team headed by William Garity in the early 1930s. It was also known as the "super cartoon camera".

The storyboard was yet another successful creation in animation technology. It is used to recheck the story and utilizes pencil sketches to review motion.

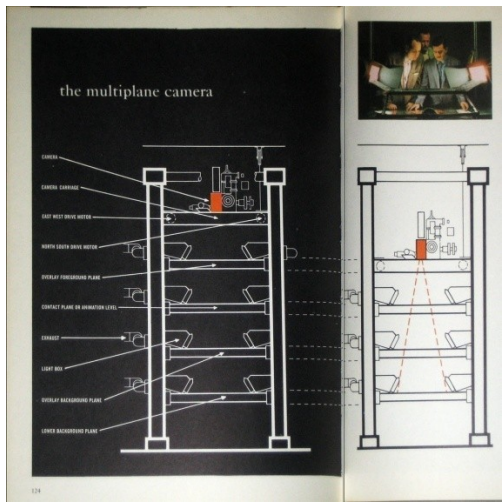 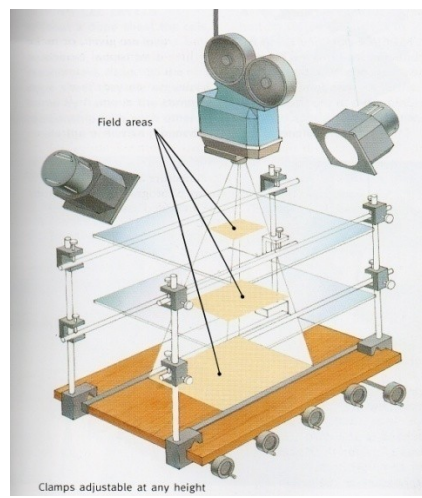

Figure 12. Multiplane camera.  Figure 13. Multiplane camera II.

Next we will discuss the first ever animated films together with some of the most famous and successful animation characters. Outstanding works in stop motion and clay motion are also elucidated, followed by a discussion of computer graphics and computer animation, i.e., 3D animation.

- **Humorous Phases of Funny Faces**

After the invention of the above-mentioned devices, J. Stuart Blackton made the first animated film in 1906. The film was entitled *Humorous Phases of Funny Faces*, and with this he became known as the father of animation. He was using a blackboard as his workplace together with chalk and an eraser as his main tools. He was able to record the animation using the "draw-stop-film-erase" method.

- **The Birth of Cartoon Characters**

The creation of the first ever animated film also inspired many animators to create their own animations. For instance, Winsor McCay drew *Gertie, the trained dinosaur* (Figure 8). It was an animated film astonishingly consisting of 10,000 drawings. The animation was shown as a film in theatres as well as at a multimedia event on stage with McCay interacting with the animated Gertie. Next in line was *Felix the Cat* (Figure 9). During the early 1920s, he became the most famous animated character. Then who could forget *Mickey Mouse?* (Figure 10). Mickey Mouse was created on November 18, 1928 and with his creation came the first successful sound animated film. *Mickey Mouse* was originally known as *Steamboat Willie* (Figure 11). He became an international star and made way for the launch of Disney Studios. Lastly, *Looney Tunes* was introduced in 1930 by Hugh Harman and Rudolp Ising run by the Warner Bros. Company. Bugs Bunny, Daffy Duck, Tweety Bird and Silvester are just a few of the main characters in this cartoon.

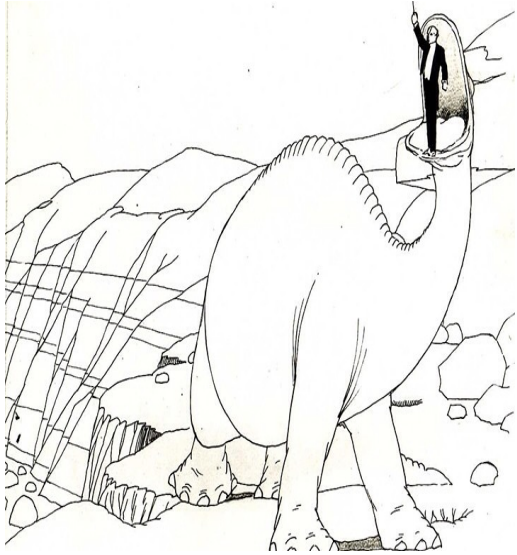
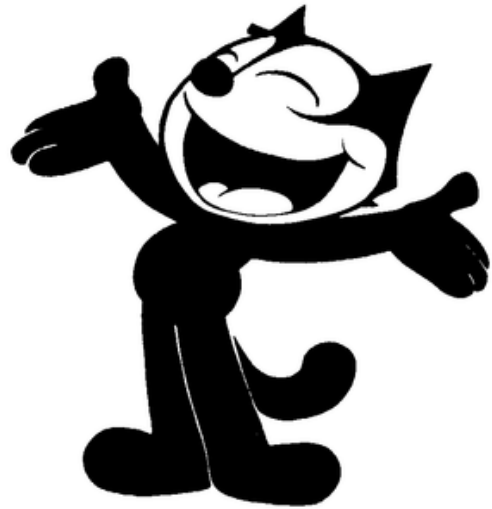

Figure 8. Gertie the trained dinosaur.    Figure 9. Felix the Cat.

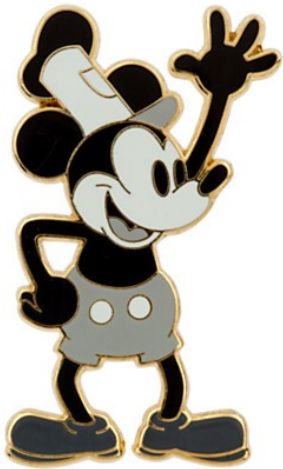
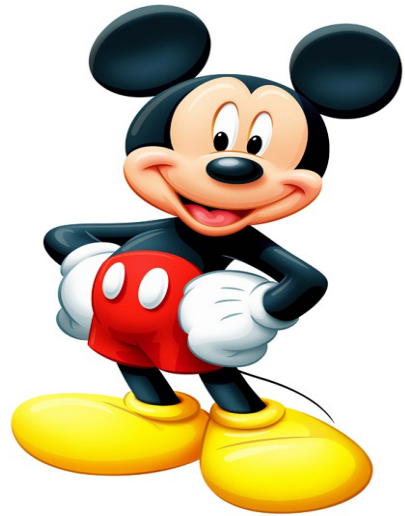

Figure 10. Steamboat Willie.    Figure 11. Mickey Mouse.

- **Stop Motion and Claymation**

Stop motion animation is used to animate things that are smaller than life size. Willis Harold O'Brian pioneered motion picture special effects, which were perfected in stop motion. He became famous after his successful work on *King Kong* (Figure 14)*,* claiming the title Dean of Stop-action Animation. Ray Harryhausen followed in the footsteps of O'Brian and became one of the most outstanding stop motion film makers through his films *Mighty Joe Young* and *The Lost World* (Figure 15).

On the other hand, claymation also became a trend. Technically, it is the art of moulding clay figures and making them move, dance, talk, sing and whatever you can think of. Frames are run together to produce the animation. *Chicken Run* and *Wallace & Grommit* are the two most successful claymations created by Aarmand Studios of the United Kingdom.

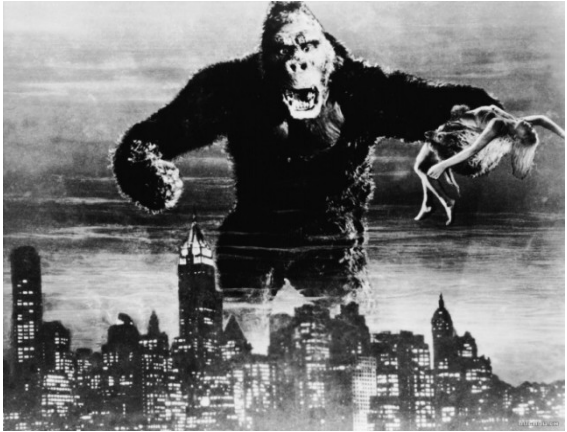 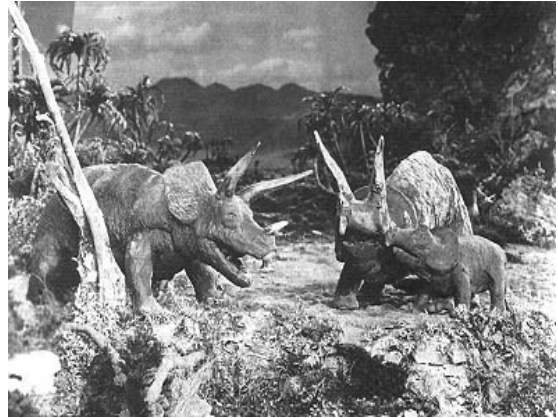

Figure 14. *King Kong*.                                   Figure 15. *Lost World*.

- **Computer Animation**

When it comes to new forms of animation, firstly let us define traditional animation - a system of animating in which the illusion of movement is presented by photographing a sequence of individual drawings on consecutive frames of film. On the other hand, computer animation is a form of pictorial presentation which refers to simulated motion pictures showing movement of drawn objects.

- **Computer Graphics and 3D Animation**

This is where graphics are created using computers and the illustration of image data by a computer particularly with the help of respective graphic hardware and software such as Superpaint. It is used to replace physical models then create realistic intermixed elements with the live action. 3D animation is today's animation. By using some sophisticated software and looking at the *Principles of Traditional Animation Applied to 3D Animation* concept, animators are able to produce outstanding and aesthetic animations such as, *Toy Story*, *Madagascar*, *Megamind*, etc.

### III. Fundamental Principles of Computer Animation

In this section, we will discuss the famous work of Pixar's John Lasseter, *Principles of Traditional Animation Applied to 3D Computer Animation*. To begin with, let us first define traditional animation: basically this is 2D animation techniques such as inbetweening (Kochanek et al., 1984; Reeves 1981), keyframe animation (Burtayk et al., 1971), multiplane background (Kevoy, 1977), scan/paint (Kevoy, 1977) and storyboarding (Gracer et al., 1970).

While 3D computer animation uses 3D models instead of 2D drawings (Lasseter, 1987). In addition to that, 3D animations were script-based with few spline-interpolated keyframe systems (Catmull, 1972). The arrival of these reliable, user-friendly, keyframe animation systems was made possible by some large companies (Lasseter, 1987) such as Abel Image Research, Alias Research Inc., and Wavefront Technologies Inc.

Nevertheless, even with such high-tech systems, most of the animations produced were all bad. One could ask, what was the problem? Their (the animators) unfamiliarity with the fundamental principles used for hand-drawn character animation for over 50 years seemed to be the reason for this.

In the late 1920s and 1930s animation was developed from an innovation to a fine art form by the Walt Disney Studios. They set up drawing classes at the Chouinara Art Institute in Los Angeles spearheaded by Don Graham. Here the students/animators learned the standardized formula of old cartoons which lead to the discovery of ways of drawing moving figures and humans (Lasseter, 1987). With this came a keen investigation of action made through the advancement of animation and its principles (Thomas et al., 1981).

There are about 11 fundamental principles of traditional animation namely: squash and stretch, timing, anticipation, staging, follow through and overlapping action, straight ahead action and pose-to-pose action, show in and out, arcs, exaggeration, secondary action and appeal.

1. **Squash and Stretch**

Technically, this pertains to the inflexibility and mass of an object by distorting its form during an action. The squashed position represents the form either flattened out by an external pressure or constricted by its own power. The stretched position constantly shows the same form in a much extended condition (Thomas et al., 1981). The most important thing to remember here is "no matter how squashed or stretched out a particular object gets, its volume remains constant" (Figure 16).

Drawing a bouncing ball is a standard animation test for the new learners in this area. The work is to complete a ball using a simple circle, and then have it fall, hit the ground and bounce back into the air (Thomas et al., 1981). There are times when some objects need not deform while doing squash and stretch. For example, a hinged figure like Luxo Jr. (Pixar, 1986), squashes by bending over on itself, and stretches by extending out completely (Figure 17). The squash and stretch principle is one of the most important principles to consider among these principles.

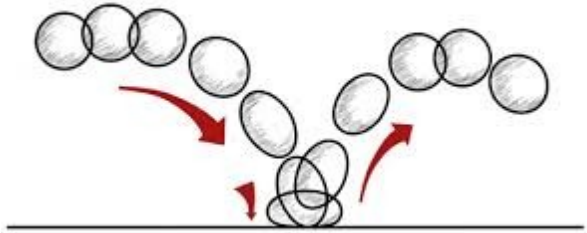 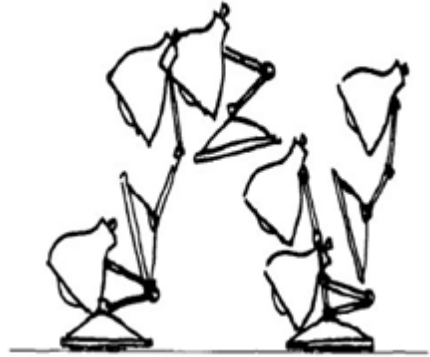

Figure 16. Squash and stretch unchanged volume.

Figure 17. Luxo Jr.

**2. Timing**

Timing is the gapping of actions to describe the weight and size of figures and the personality of characters. It marks to the viewers how well the idea behind the action is thought out. Accurate timing is critical to making ideas clear. The anticipation of an action, the action itself, and the feedback to the action are essentials, and adequate time is needed to prepare the audience for this. The audience will be confused if too much time is spent on any of this, and equally if not enough time is spent, the motion may be finished before the audience becomes aware of it, consequently wasting the idea (Whitaker et al., 1986).

**3. Anticipation**

The preparation for the action, the action proper, and the termination of the action are the main parts of anticipation. It is also a tool to gain the audience's attention, to prepare them for the next movement and to guide them to guess it before it actually happens. Anticipation is usually used to clarify what the next action is going to be. For example, when a character tries to grab an object, firstly he extends his arms as he looks at the object and makes facial expressions to indicate that he is going to do something with that particular object. The anticipatory moves might not reveal why he is doing something, however, the "what is he going to do next?" question is critical (Thomas et al., 1981).

More examples of anticipation include:

- A heavy object is present and a particular character is trying to pick it up. Bending right down before picking up the object, aids the momentum of the character to lift heavy things - this is known as exaggerated anticipation.
- Imagine a fat character stuck in the seated position. For the character to stand up, he would need to bend his upper body forward, with his hand on the armrests of the chair for support, before pushing his arms and using the momentum of his body (White et al., 1986).

4. **Staging**

This refers to the arrangement of an idea entirely and clearly, and it originates from 2D hand-drawn animation. For better understanding, action is staged. For easy recognition, personality is staged. To affect the mood of the audience, expressions and moods are staged (Thomas et al., 1981). In presenting animation, actions should take place one at a time. If there is lot of movement at the same time, the audience will get lost and will not know where to focus, thus the learning is affected negatively. Each and every action and idea should be staged in the strongest and the simplest way before moving on to the next phase.

5. **Follow Through and Overlapping Action**

Since anticipation is the preparation of an action, termination of an action is defined as follow through. Actions do not often come to a sudden and full stop, but are usually carried past their ending point. For example, after releasing a thrown ball, the hand continues past the real point of release.

Regarding the movement of any particular objects, the actions of the parts are not synchronized; there should be an initiation for the movement, e.g., the engine of a train is also known as the lead. Let us consider walking where the action starts with the hips. As the walking starts, the hips swings forward, which then sets the legs into motion. The hip is the lead, next the legs follow then the torso, the shoulders, arms, wrists and finally the fingers all come into play.

6. **Straight Ahead Action and Pose-to-pose Action**

There are two main approaches to hand-drawn animation namely: straight ahead action and pose-to-pose action. The main difference between these two is the readiness. Straight ahead action starts from nothing and hopes to continue as ideas pop-out from the mind. On the other hand, pose-to-pose action makes sure that everything is ready, that the characters are drawn and prepared, and that it is just a matter of connecting them in size and action with one another, and then the inbetweens are drawn.

7. **Slow In and Out**

The gapping of the inbetween drawings between the extreme poses describes the slow in and out. In mathematics, this is defined as the second- and third-order continuity of motion. The "slowing out" of one pose then the "slowing in" to the next basically describes the logic behind slow in an out. In essence it pertains to the timing of the inbetweens (Figure 18).

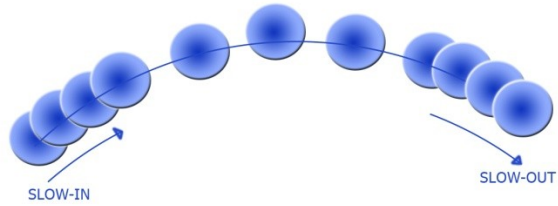

Figure 19. Slow in and slow out.

In addition, the animator should point out the position of the inbetweens with a "timing chart" as shown below (Figure 20). With this, animation becomes more alive and spirited.

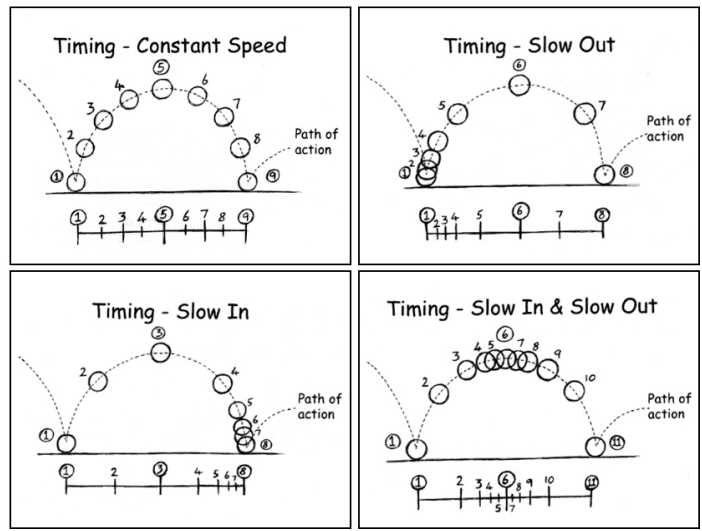

Figure 20. Timing chart.

## 8. Arcs

An arc describes the visual path of action from one previous extreme to another. It is also the most economical routes by which an object, shape or form can be in motion from one place to another. In order to make the animation smoother and less stiff the use of arcs is highly advisable as opposed to just using a straight line. In some cases, even an arc can turn into a straight path, for instance, the free fall in physics. The use of arc in a free fall animation would be illogical. Generally, the path of action from one extreme to another is maintained by a similar spline in 3D keyframe computer animation systems and that which controls the timing of the inbetween values.

## 9. Exaggeration

The distortion of the shape of an object or making an action look more violent or unrealistic does not necessarily mean exaggeration. It is the animator's job to go to the heart of anything or any idea and develop its real meaning, comprehending the cause for it, so that the viewers will also understand it. For instance, a sad character could be sadder, or if he is bright, make him dazzling.

**10. Secondary Action**

Secondary action is basically the action that comes directly from another action. Secondary actions are essentials in building interest and including a realistic complexity to the animation. From the name itself, secondary action is always inferior to the primary action. If it conflicts, becomes more interesting, or dominates in any way, it is either the wrong choice or is staged improperly (Thomas et al., 1981). To give you a simple example, if the character's movement of the body is the main idea, the facial expression becomes the secondary action (Thomas et al., 1981).

**11. Appeal**

Appeal is mainly composed of the quality of charm, pleasing design, plainness, communication and magnetism. Appeal is what the person or the audience likes to see. A figure or object that has appeal will easily catch the audience's eye and holds it while the audience appreciates the object. An inadequate drawing or design lacks appeal. Any complicated design with clumsy shapes and strange moves is hard to understand and likewise lacks appeal.

After all these principles *personality* comes into the picture. Personality refers to the wise application of all the principles of animation. Notice that when the character animation is successful the audience is entertained. This is because the accomplishment of character animation depends in the personality of the characters regardless of whether it is drawn by hand or computer. The story also becomes more essential and evident to the audience if good personality is shown. Moreover, in personality there is no one type of character created. Take a boy playing a ball and his dad into account, their physical characters are very different so too are the emotions. The boy is very lively and active while the father is just calm.

**IV. Using Computer Animation in Education**

Nowadays, we live in a digital era: it is inescapable. The advancement of technology and the drastic changes in the surroundings affects our needs and desires, be it psychologically, socially or emotionally.

Similarly, the need for change in education as time has passed has strongly accelerated. These days, the old chalk and talk methods in front of a blackboard, or even the more updated whiteboard, marker and projector, are simply not enough for effective teaching and learning. As Abbas (2012) comments, "students tire of this teacher-centered model…..and complain that the…..class is very boring and monotonous and they want something new and different". Though the old methods may still be present, increasingly there is a demand for a more competitive tool which will supply the needs of the students more effectively. This involves "changes in both the instructional strategy and also the teaching and learning environment" (Vonganusith & Pagram, 2008).

Computer animation, specifically educational computer animation, can be considered as one of the main tools available for teachers to use to promote effective learning nowadays. In this part, we will enumerate the usage of computer animation in education:

1. To help the learners to visualize something which can't be seen easily in the real world (Ainsworth, 2008).

    a) e.g. the movement of atoms in a gas (Russell et al., 2000) shown in Figure 21.

    b) the shifting movements of the continents (Sangin et al., 2006) shown in Figure 22.

2. To illustrate events that are not inherently visual (Ainsworth, 2008). Animation clarifies relationships through visual means (Weiss, 2002).

    a) e.g. computer algorithms (Kehoe, Stasko, & Taylor, 2001).

    b) e.g. stages in a mathematical solution (Scheiter, Gerjets, & Catrambone, 2006).

3. To serve a decorative or cosmetic function (Levin, Anglin & Carney, 1987; Weiss, 2002).

    a) e.g. special animated effects sometimes can dazzle and impress students in the opening of a lesson (Weiss, 2002).

4. To gain attention (Rieber, 1990a).

    a) e.g. interesting special effects for transitions between instructional frames, screen washes, moving symbols or characters, and animated prompts (Weiss, 2002).

5. To provide feedback.

    a) e.g. a dancing bear, a unicycle-riding clown, or fireworks used as feedback can motivate learners to strive for correct answers (Weiss, 2002), though according to Surber and Leeder (1998) overuse of colourful graphics does not enhance motivation.

6. To use it as part of the presentation strategy (Weiss, 2002). This is particularly helpful when presenting highly abstract or dynamic processes (DiSessa, 1982; Kaiser, Proffett, & Anderson, 1985; Rieber, 1990a, 1991).

    a) e.g. animation might be particularly useful in helping students understand the flow of blood through the body (Weiss, 2002).

    b) e.g. the inner life of the cell.

7. To assist the users with animated agents (Johnson, Rickel & Lester, 2000), where lifelike characters are animated to include gesture and movement (Ainsworth, 2008) shown in Figure 23.

8. To assist with language teaching and learning (Bikchentaeva & Ziatdinov, 2012). The use of computer animation in language development can be justified according to several theoretical paradigms: constructivism, which places emphasis on a learner's active engagement with the learning process in order to construct meaning out of the available input (for instance, Williams & Burden, 1997); cognitive psychology, which acknowledges a learner's active attempts to understand, acquire, store and use knowledge (for instance, Skehan' 1998); the Affective Filter Hypothesis (Krashen & Terrell, 1983) which recognizes the importance of emotional factors such as motivation, interest and anxiety in effective learning; the Noticing Hypothesis, which suggests that learning takes place most effectively when the material to be learnt attracts and holds the learner's attention (Schmidt, 1995); learning style, which suggests that individuals have different preferred ways of learning and that they learn best when their particular style is accommodated (for instance, Reid, 1987); and learning strategies, according to which different activities are more or less effective for the promotion of learning in different students (for instance, Griffiths, 2013; Oxford, 2011).

Computer animation can be used in all areas of language cognition and skills development:.e.g. expanding vocabulary. Computer animation could be particularly useful when teaching verbs (for instance, run. jump. kick, climb, fall etc) where the action dimension can be especially difficult to convey using conventional static teaching methods.

a) e.g. explaining grammar. Computer animation techniques can be specially effective, for instance, with drilling new grammar until it becomes automatic, a process which can be extremely monotonous and demotivating using traditional pedagogical techniques.

b) e.g. modeling pronunciation. According to Massaro (2003, p.172) "combined with principles from linguistics, psychology and pedagogy, (computer animation) technology has the potential to help individuals…..learning a new language".

c) e.g. providing listening input. It is not always easy to obtain an adequate supply of material to develop listening skills, but computer animation programmes have the potential to provide an unlimited source of such material.

d) e.g. stimulating speaking practice. By means of interactive programmes, it is possible to develop students' speaking skills in an interesting, non-threatening environment

e) e.g. presenting interesting reading material. An unlimited source of reading texts, matched to the students' interests and ability levels, can be provided by means of animated computer programmes.

f) e.g. motivating students to write. Receptive skills (listening, reading) can be used to stimulate productive skills (speaking, writing) by means of the attention generated from computer animation.

The use of computer animation in education has broadened and continues to grow at a rapid pace. Because of the speed of change, we need to train teachers who are capable of dealing with technology. Also, as Abbas (2012) explains, teachers need to learn how to adopt new roles, such as those of facilitator and guide, integrator, researcher, designer and collaborator. Since the new pedagogical paradigms involve more than the mere transmission of knowledge, which was once considered the norm, teachers may need training in how to develop the professional expertise which using computer animation in education requires.

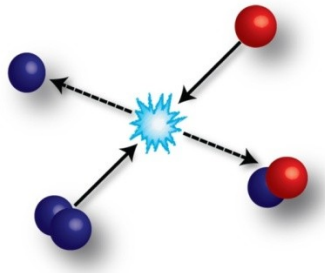 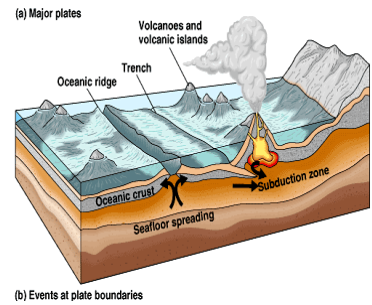

Figure 21. Movement of atom gas.     Figure 22. Illustration on plate movement.

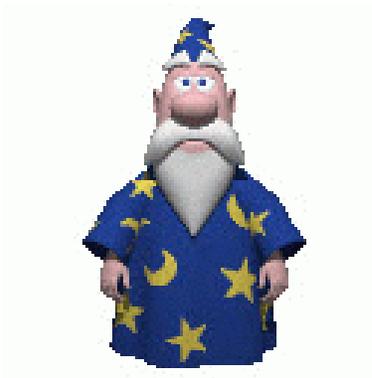 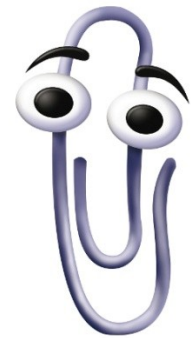

Figure 23. Microsoft's famous animated agents Merlin(L) and Clipit(R).

## V. Future of Computer Animation

In order to effectively use animation it is useful to understand how people learn from pictorial and verbal media (Mayer and Moreno, 2002). In order to promote the cognitive processes required for meaningful learning such as selecting, organizing and integrating, multimedia presentations should be designed in ways to promote just that (Mayer and Moreno, 2002).

In the next millennium, pictorial forms of teaching are likely to expand as a complement to verbal forms of teaching (Pailliotet and Mosenthal, 2000). As stated by Mayer and Moreno (2002), "animation is a potentially powerful tool for multimedia designers, but its use should be based on cognitive theory and empirical research… The future of instructional animations is bright to the extent that its use is guided by cognitive theory and research."

## VI. Conclusions and Future Direction

Animation, which is basically a form of pictorial presentation, has become the most prominent feature of technology-based learning environments. Animation refers to simulated motion pictures showing movement of drawn objects. Educational animation is one of the most elegant tools for presenting materials for learners. Its significance in helping learners to understand and remember information has greatly increased since the advent of powerful graphics-oriented computers. It may be very useful for learning about some topics in the natural sciences, where educational modelling and preparing materials convenient for learning can reduce the time required in class and increase the efficiency of the educational process. On the other hand, English language and literature can benefit from the possibility of creating animated multimedia books. By utilizing animation, students develop skill competencies in visual communication, storytelling, observation and sensory aspects, problem-solving and innovative aspects, e.g., concentration, as well as cognition, ethics and aesthetics.

Our future work aims to establish an inter-disciplinary field of research looking into greater educational effectiveness. With today's high educational demands, traditional educational methods have shown deficiencies in keeping up with the drastic changes observed in the digital era. Without taking into account many significant factors, educational animation materials may turn out to be insufficient for learners or fail to meet their needs. However, the applications of animation and ergonomics to education have been given inadequate attention, and students' different temperaments (sanguine, choleric, melancholic, and phlegmatic, etc.) have not been taken into account. We suggest there is an interesting relationship here.

We shall propose essential factors in creating educational animations. The fundamentals of design are divided into design principles (alignment, balance, emphasis, unity, proximity, rhythm) and design elements (colour, line, shape, space, texture, value). Using these fundamentals and forming sequences of images to create an illusion of aesthetic movement, i.e., movement of objects along aesthetic paths drawn by so-called fair curves (class A Bézier curves, pseudospirals, superspirals, quadratic Bézier curves with monotonic curvature function, Pythagorean hodograph spirals), we intend to create educational materials considering different types of student temperament and preferences. To give an ideal model of educational animations, in this study we use design and animation software such as CorelDraw Graphics Suite X6 and ToonBoom. CorelDraw Graphics Suite X6 is broad software for graphic design, page layout and picture editing. ToonBoom is the leading international developer of digital content and animation software solutions, providing the most cohesive and state-of-the-art toolsets available today. It serves educators at all levels searching for ways to integrate an art-based curriculum into their existing classroom environments. Educational materials designed using the above software increase literacy rates and facilitate the achievement of overall learning objectives.

We believe that this study is likely to have wide benefits in the field of education. Designing educational materials with the aid of the mentioned software, while considering the types of temperament of students, is a really promising avenue to improve the learning process. Teachers will be able to feel more confident in the presentation of their lessons, in addition, they will become more competitive and professional.

## VII. Acknowledgement

This work is supported by the Scientific Research Fund of Fatih University under the project number P55011301_Y (3141).

[34] Mayer, R. E., Hegarty, M., Mayer, S., & Campbell, J. (2005). When static media promote active learning: Annotated illustrations versus narrated animations in multimedia instructions. Journal of Educational Psychology: Applied, 11, 256-265.

[35] Mayer, R., & Anderson, R. (1992). The instructive animation: helping students build connections between words and pictures in multimedia learning. Journal of Educational Psychology, 84(4), 444-452.

[36] Mayer, R.E. (1996). Learners as information processors: Legacies and Limitations of educational psychology's second metaphor. Educ. Psychol., 31, 151-161.

[37] Mayer, R.E. (1999). Multimedia aids to problem-solving transfer. Int. J. Educ. Res. 31, 661-624.

[38] Milheim, W.D. (1993). How to use animation in computer assisted leaning. British Journal of Educational Technology, 24(3), 171-178.

[39] Moreno, R., & Valdez, A. (2005). Cognitive principles of multimedia learning: The role of modality and contiguity, Journal of Educational Psychology, 91, 358-368.

[40] Moreno, R., and Mayer, R. E. (2000a). A coherence effect in multimedia learning: The case for minimizing irrelevant sounds in the design of multimedia instructional messages. J. Educ. Psychol. 92: 117-125

[41] Moreno, R., and Mayer, R. E. (2000b). Engaging students in active learning: The case for personalized multimedia messages. J. Educ. Psychol. 93: 724-733.

[42] Moreno, R., and Mayer, R. E. (2002). Animation as an aid to multimedia learning. Educational Psychology Review. 14.1: 87-98. Santa Barbara, California.

[43] Moreno, R., Mayer, R.E., and Lester, J.C. (2000). Life-like pedagogical agents in constructivist multimedia environments: Cognitive consequences of their interaction, In Proceedings of ED-MEDIA 2000, AACE Press, Charlottesville, VA, pp. 741-746.

[44] Oxford, R. (2011). Teaching and Researching Language Learning Strategies. Harlow: Pearson Longman.

[45] Pailliotet, A. W., and Mosenthal, P.B. (eds). (2000). Reconceptualizing Literacy in the Age of Media, Multimedia, and Hypermedia, JAI/Ablex, Norwood, NJ, USA.

[46] Paivio, A. (1986). Mental Representations: A Dual Coding Approach, Oxford University Press, Oxford, England.

[47] Pane, J. F., Corbett, A. T., & John, B. E. (1996). Assessing dynamics in computer-based instruction. Proceedings of ACM CHI'96 Conference on Human Factors in Computing Systems, Vancouver, Canada.

[48] Pass, F., Renkl, A., & Sweller, J. (2003). Cognitive load theory and instructional design: Recent developments, Educational Psychologist, 38, 1-4.

[49] Plötzner, R., & Lowe, R. (2004). Dynamic visualisations and learning, Learning and Instruction, 14, 235-240.

[50] Price, S.J. (2002). Diagram Representation: The cognitive Basis for Understanding Animation in Education (Technical Report CSRP 553): School of Computing and Cognitive Sciences, University of Sussex, UK.

[51] R.E. Weiss et al. (202). Computers in Human Behavior, 18, 465-477.

[52] Reeves, W. (1981). Inbetweening for Computer Animation Utilizing Moving Point Constraints, SIGGRAPH '81, Computer Graphics, Vol. 15, No. 3, pp.263-270.

[53] Reid, J. (1987). The Learning Style Preferences of ESL Students. TESOL Quarterly, 21/1, 87-111.
22

**Authors**

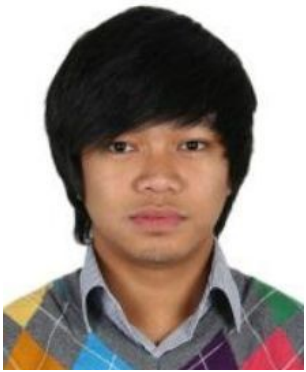

**Sajid Musa** is a bachelor student in the Department of Computer & Instructional Technologies at Fatih University, Istanbul, Turkey. He has graduated with a silver medal from Fountain International School, San Juan, Philippines in 2011. Despite his youthfulness, he has managed to publish several manuscripts on algorithmic thinking and educational systems in the Philippines. His current research interests include instructional design, informational technologies in education, rapid mental computation systems, and computer and educational animation. In the future he wishes to continue his research at Harvard University, USA.

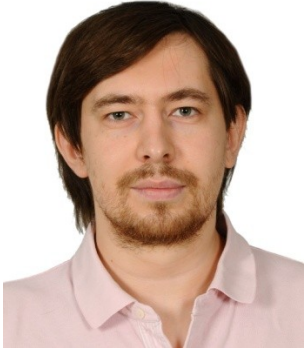

**Rushan Ziatdinov** is an assistant professor in the Department of Computer & Instructional Technologies at Fatih University, Istanbul, Turkey. In the beginning of his scientific career he held the positions of assistant professor in the Department of Geometry & Mathematical Modelling at Tatar State University of Humanities and Education and in the Department of Special Mathematics at Tupolev Kazan National Research Technical University (Kazan University of Aviation), Kazan, Russia. Then he moved to Seoul National University, South Korea, where he was a postdoc in the Computer-Aided Design and Information Technology Lab. His current research interests include geometric modelling, computer-aided geometric design, educational animation, computer graphics in mathematics education, the use of computer models in natural science education, instructional technologies in mathematics education, computer modelling, realistic modelling, computer-aided aesthetic design and educational ergonomics.

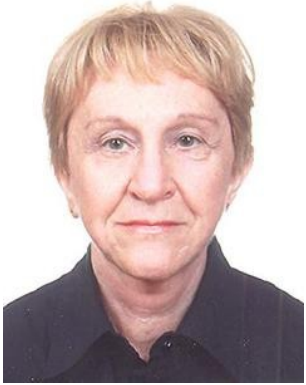

**Carol Griffiths** has many years of experience as a teacher, manager and teacher trainer in the field of English Language Teaching. She completed a PhD researching language learning strategies at the University of Auckland, New Zealand, and learner issues continue to be her main research interest. Carol is currently working as an associate professor at Fatih University in Istanbul, Turkey, having previously worked in New Zealand, Indonesia, Japan, China, North Korea and the UK.